\documentstyle[preprint,prl,epsf,aps]{revtex}
\newcommand{\nix}[1]{}
\begin{document}

\title{Spin sensitive bleaching
and monopolar spin orientation
in quantum wells}

\author{S.D. Ganichev$^{1,2}$, S.N. Danilov$^1$, V.V. Bel'kov$^2$,
E.L. Ivchenko$^2$,\\
M.~Bichler$^3$, W. Wegscheider$^{1,3}$, D.~Weiss$^1$, and W. Prettl$^1$
}

\address{$^1$ Institut f\"{u}r Experimentelle und Angewandte Physik,\\
Universit\"{a}t Regensburg,
D-93040 Regensburg, Germany\\
$^2$ Ioffe Physico-Technical Institute, RAS,
194021, St. Petersburg, Russia\\
$^3$ Walter Schottky Institut, TU M\"{u}nchen, 85748, Garching, Germany
}

\date{\today}
\draft
\maketitle
\begin{abstract}

Spin sensitive bleaching of the absorption of far-infrared radiation
has been observed in $p$-type GaAs/AlGaAs
quantum well structures. The absorption of circularly polarized radiation
saturates at lower intensities than that of linearly polarized light due to
monopolar spin orientation in the first heavy hole subband.
Spin relaxation times of holes in $p$-type material in the range of tens of ps
were derived from the intensity dependence of the absorption.

\end{abstract}


\newpage

A substantial portion of current research in
condensed-matter physics is directed towards understanding
various manifestations of spin-dependent phenomena like
giant magneto-resistance, heavy fermions, Kondo scattering,
superconductivity and others. In particular, the spin of electrons and
holes in solid state systems is the decisive ingredient for
active spintronic devices\cite{Datta,Prinz} and several
schemes of quantum computation\cite{Kane,Loss,Divincenzo00}.
Especially the combination of ferromagnetic materials with
semiconductors seems to be a
promising combination for novel functional concepts. Open problems
which have to be addressed
in this respect involve spin-injection into semiconductors,
spin relaxation in low dimensional
semiconductor structures as well as spin detection. Significant
progress was made recently: it was
shown that spin polarized electrons (or holes) can be injected from semimagnetic (or ferromagnetic)
semiconductor materials into semiconductors\cite{Fiederling,Ohno}. The presence of spin
polarized electrons can be probed by analyzing the Kerr effect~\cite{Malajovich} or by analyzing the
degree of circular polarization of light which gets emitted when polarized electrons (holes)
recombine with holes (electrons). The inverse process, exciting free carriers
by circularly
polarized light due to optical orientation~\cite{Meier} is frequently used to prepare an ensemble of
spin polarized carriers. In low-dimensional systems with band splitting in 
k-space due to k-linear terms in the Hamiltonian 
optical excitation  not only leads to a spin polarized ensemble of electrons but
also to a current whose sign and magnitude depends on the degree of circular
polarization of the incident light (circular photogalvanic effect,\cite{PRL01}).

For the realization of spintronic devices
sufficiently long spin dephasing times  in semiconductor
quantum well (QW) structures are crucially needed. Spin transport must
occur without destroying the relevant spin information.
Current investigations of the spin lifetime in semiconductor
devices~\cite{Awschalom1,Oestreich,Awschalom2,Shah,Sham,Fabian} are based on
optical spin orientation by interband excitation and further  tracing
the kinetics of polarized photoluminescence and
investigation of the dynamics of subsequent relaxation of spin
polarized electron-hole pairs. Studies of bi-polar spin
orientation, where both
electrons and holes  got excited,
gave important insights into  the
mechanisms of spin relaxation of photoexcited free carriers.
 We show below that by combining  the circular photogalvanic
effect (CPGE)~\cite{PRL01} with
saturation (bleaching of absorption)
spectroscopy
\cite{Beregulin82,Pigeon83,Beregulin87,Helm93,Li94}
we are able  to probe  spin relaxation for monopolar
spin orientation. In contrast to the conventional methods of
optical spin orientation, in our measurements only one type
of charge carriers (electrons or holes) gets spin oriented
and is involved in relaxation processes. This is achieved by using
radiation in the terahertz range which excites intraband or intersubband,
but no interband (from valence to conduction band), transitions.
Monopolar spin
orientation allows to study
spin relaxation
without electron-hole interaction and exciton
formation. The most
important advantage of monopolar spin orientation is that
relaxation processes can be investigated for electrons in
n-type material and for holes in p-type material;
both can be measured independently.
These
conditions have not been met previously in quantum
well structures where, due to interband excitation,
only the spin relaxation times of optically generated minority
carriers were accessible (for reviews see~\cite{Shah,Sham,Fabian}).

Here we report  the first observation of spin sensitive
bleaching of the
heavy hole $hh1$- light hole $lh1$ absorption  in $p$-type  QW
structures which allows  to investigate spin relaxation for a monopolar
spin orientation.
The basic physics is sketched in Fig. 1. Exciting with circularly polarized light results in direct
intersubband transitions (solid arrow) which depopulate and populate selectively spin states in the
valence subbands ($hh1$ and $lh1$) and cause a monopolar spin polarization. Spin
relaxation inside the $hh1$-subband is characterized by the relaxation time $\tau_s$. Relaxation from the
$lh1$- subband back to $hh1$  is characterized by a spin independent energy relaxation time $\tau_e$ as the
dominating transitions in this energy range are indirect (broken arrows in Fig. 1), mediated by
phonons. The absorption coefficient $\alpha$ is proportional to the difference of the populations of the initial
and final states. At high intensities the absorption coefficient decreases since the photoexcitation rate
becomes comparable non-radiative relaxation rate into the initial state.
Thus absorption bleaching of circularly polarized radiation governed
by both the hole spin relaxation in the initial state
and the energy relaxation
of photoexcited carriers characterized
by the spin relaxation time $\tau_s$ and the energy
relaxation time $\tau_e$, respectively.
In contrast to circularly polarized light,
optical transitions induced by  linearly polarized light are not
spin selective and the saturation is controlled by the energy relaxation
of photoexcited carriers only. The difference in absorption
bleaching for circularly and linearly polarized
radiation can be observed experimentally as is pointed out below.
The method introduced here can
also be applied for $n$-type QWs
using direct inter-subband transitions in the conduction band.

The experiments have been carried out on 
modulated doped p-GaAs/AlGaAs
(311)-MBE-grown samples with a single QW or 20 QWs
of the width $L_W$=15 nm and with the 
period $d$= 1300~nm for  multiple QW-structures.
Samples with free carrier densities $p_s$  of 
about $2\cdot 10^{11}$ cm$^{-2}$ and
mobilities
around 5x10$^5$~cm$^2$/Vs were studied in the range from liquid
helium  to room
temperature. A pair of ohmic contacts have been centered
on opposite sample edges along the direction $x \parallel
[1\bar{1}0]$ (see inset in Fig.~2)~\cite{PRL01}.
As  terahertz radiation source a high power
far-infrared molecular laser, optically pumped by a
TEA-CO$_2$ laser, has been used delivering 100~ns pulses with
intensities up to 1~MW/cm$^2$ at wavelength $\lambda
=$148~$\mu$m. The radiation induces direct optical
transitions between the first heavy-hole  and the first
light-hole valence subband (see Fig.~1). 
A crystalline quartz $\lambda/4$ plate has been used
to obtain from the initially linearly polarized  laser light circularly
polarized radiation with the degree of  circular polarization $P_{circ}$   
equal to $\pm 1$ for right and left handed circularly polarized light, respectively.

The weak absorption of terahertz radiation
by free carriers in
QWs is difficult to determine by direct  transmission measurements
 particularly in the case of bleaching at high power levels.
Therefore the  nonlinear behavior of the absorption
has been investigated employing
the recently observed circular and linear photogalvanic
effects \cite{PRL01,APL00}. Both, the circular photogalvanic effect
(CPGE) and the linear photogalvanic effect (LPGE)
yield an electric current
in $x$ direction\cite{photogalvanic}.
The absorption coefficient is proportional to 
photogalvanic current $j_x$
normalized by the 
radiation intensity $I$~\cite{book}. By choosing circularly
or
linearly polarized radiation we thus
obtain a photoresponse corresponding to the absorption coefficient of circularly
or linearly polarized radiation, respectively.

Our measurements, displayed in Fig.~2, indicate that the photocurrent $j_x$
depends at low power levels linearly on the
light intensity and gradually saturates
with increasing intensity $I$ as $j_x\propto I/(1+I/I_s)$, where
$I_s$ is the
saturation intensity. This behavior of the current corresponds to a
constant absorption
coefficient at low power levels and decreasing absorption with rising
intensity.
Saturation intensities $I_s$ have been
measured for a wide range of temperatures between 4.2 K and
200 K. The key experimental result is plotted in Fig.~3
and show  that the magnitude $I_s$ for the
photogalvanic response on circularly polarized radiation
is generally smaller than that for
linearly polarized radiation. The experimentally
obtained values increase from about 10 kW/cm$^2$ at liquid
helium temperature to 300 kW/cm$^2$ at 200~K.
At room temperature
the saturation intensities get
non-measurably large.

Over the whole temperature range the  holes occupy in equilibrium
the lowest heavy-hole subband $hh1$.
Absorption of far-infrared radiation occurs by direct optical transitions
from $hh1$ to the first light hole subband $lh1$ and for 
the  wavelength 148~$\mu$m (photon energy $\hbar\omega=8.3$~meV) 
used here takes place close to $k = 0$, as is sketched  in Fig.~1.
Thus, the optical dipole selection rules for the absorption are $\Delta m = \pm 1$ with
angular momentum quantum number $m = \pm 3/2$ for the initial state and
 $m = \pm 1/2$ for the final state\cite{Ferreira}.
The insets in Fig.~3 show the corresponding transitions for linear (top left)
and circular (bottom right) polarization by full arrows. 
Linearly polarized radiation has
been decomposed in right and left handed circularly polarized
light of identical amplitudes. Broken lines in these insets  indicate
non-radiative energy and spin relaxation transitions.

Linearly polarized radiation (top left inset in Fig.~3) equally depopulates both spin-up
and spin-down states of the heavy hole subband and populates the
first light hole subband. With rising intensity these
non-equilibrium  populations approach each other causing the
bleaching of absorption which is controlled by the energy
relaxation time $\tau_e$. In contrast to linear
polarization the absorption of circularly polarized light is
spin selective because only one type of spin component is
involved in the absorption process (illustrated in the  right bottom
inset of Fig.~3).
During energy relaxation to the initial state in subband $hh1$ the
holes loose the photoinduced orientation due to rapid relaxation~\cite{Ferreira}. 
Thus, spin orientation occurs
in the initial
subband $hh1$, only. Bleaching of absorption is hence controlled by two time
 constants, the  energy relaxation time $\tau_e$ and the
spin relaxation time $\tau_s$. Note that $\tau_e$ is the same for circular and 
linear polarization.
If $\tau_s$ is of the order of
$\tau_e$ or larger, bleaching of absorption
becomes spin sensitive and the saturation intensity of
circularly polarized radiation drops below the value of linear
polarization.

Spin sensitive bleaching can be analyzed in terms of
excitation-relaxation kinetics
taking into account both optical excitation and
non-radiative relaxation processes. The
probability  rates for direct optical transitions from
the $hh1$ states with $m=\pm 3/2$ to higher subbands are
denoted as $W_{\pm}$. For linearly polarized light, $W_+$
and $W_-$ are equal. For the circular polarization, right
handed, $\sigma_+$, or left handed, $\sigma_-$, the rates
$W_{\pm}$ are different but, due to  time inversion
symmetry,  satisfy the condition $W_{+} (\sigma_{\pm})
= W_{-} (\sigma_{\mp})$. If $p_+$ and $p_-$ are
the 2D densities of heavy holes with  spin $+ 3/2$ and $- 3/2$,
respectively, then the rate equation for $p_+$ can be
written as
\begin{equation} \label{rate}
\frac{\partial p_+}{\partial t} + \frac{p_+ - p_-}{2 \tau_s} =
-W_+ + \frac12 (W_+ + W_-).
\end{equation}
The
corresponding equation for $p_-$ is obtained by exchange of
indices $\pm \rightarrow \mp$. Since the laser pulse duration  
was longer than any expected relaxation time we
consider the steady-state solution of the  rate equations and
omit the time derivative in Eq.~(\ref{rate}).
The
second term on the left-hand side of Eq.~(\ref{rate})
describes the spin relaxation trying to equalize the
polarization of  the $\pm 3/2$ states.
The first term on the right-hand side
describes the removal of holes from the $hh1$ subband due to photoexcitation
while the second term characterizes the relaxation of holes which come down
to the $+ 3/2$ and $- 3/2$ states with equal rates (see insets in Fig.~3). The right side 
of the Eq.~(\ref{rate}) is proportional 
to $W_+ - W_- = \frac{\alpha  d I}{\hbar \omega} (\rho_0 P_{circ} - \eta \rho)$,
where  $\rho = (p_+ - p_-)/p_s$ is the hole spin polarization degree,
$\rho_0$ is the excitation induced spin polarization, and
$\eta \approx f_i/(f_i-f_f) \approx 1$ describes the difference between the population 
of the initial state, $f_i$, and the final state, $f_f$, respectively.

Bleaching of absorption with  increasing intensity is
described by
the function $\alpha = \alpha_0 [1 + ( I / I_{se}) ]^{-1}$
where $\alpha_0$
is the absorption coefficient at low intensities and $I_{se}$ is the
characteristic
saturation intensity controlled by energy relaxation of the hole gas. Since the
photocurrent $j_{LPGE}$ induced by the linearly polarized radiation
is proportional
to $\alpha I$, one has
\begin{equation}
\frac{j_{LPGE}}{I} \propto \frac{1}{1 + \frac{I}{I_{se}}}\:.
\end{equation}
The photocurrent
$j_{CPGE}$ induced by the circularly polarized radiation is
proportional to  $W_+ - W_-$. Solving Eqs. (1,2) in
the steady-state
regime we obtain
\begin{equation}
\frac{j_{CPGE}}{I} \propto \frac{1}{1 + I \left( \frac{1}{I_{se}} + \frac{1}{I_{ss}}
\right) } \:,
\end{equation}
where $I_{ss}= \hbar\omega p_s /(\alpha_0 d \tau_s)$
is the saturation intensity controlled by the hole
spin relaxation.
The saturation intensities $I_{ss}$ and $I_{se}$ were
extracted  from the measured saturation intensities
$I_s$ of linear and circular photogalvanic current (Fig.~3).
Using $I_{ss}$ together with the  absorption coefficient $\alpha_0$,
calculated  after~\cite{Golub}
for the  wavelength used here, spin relaxation times $\tau_s$ have been derived.
The results are plotted in
Fig.~4 as a function of temperature.
At low temperatures the relaxation times vary like $T^{-\frac{1}{2}}$.

The magnitude of the observed hole spin relaxation time
$\tau_s$ in the first heavy hole subbband $hh1$
is in very good agreement to previously published
data obtained from the interband recombination kinetics of
circularly polarized 
photoluminescence\cite{Shah,Sham,Fabian,Damen91_2,Barad92,Baylac,Adachi,Potemski}.
These works, the relation between spin relaxation time and free carrier
density has been discussed in terms of the D'yakonov-Perel and
the Bir-Aronov-Pikus mechanisms. In our case of monopolar spin
orientation photocreated carriers and electron-hole pairs do
not exist. Thus, in contrast to all previous experiments were
 hole spin relaxation times were probed by interband
excitation, the
Bir-Aronov-Pikus mechanism is absent
and the spin relaxation is not affected by
high density photocreated carriers, exciton formation and
interband recombination.
The D'yakonov-Perel mechanism was  investigated theoretically
by Ferreira and Bastard\cite{Ferreira,Bastard2} for spin relaxation
of $hh1$-holes in GaAs based QWs.
The values of $\tau_s$ on the order of 10~ps as well as the observed temperature
dependence $\tau_s\propto T^{-1/2}$ are
in accordance with these calculations for samples with 
parameters as in our experiment. 

In conclusion, our experimental results demonstrate that
absorption of terahertz radiation by inter valence band
transitions in $p$-type QWs becomes spin sensitive at high power levels.
The saturation of circularly polarized radiation, which is
measured in QWs by the circular photogalvanic effect, yields
the spin relaxation times of majority carriers, in our case holes.
Finally we would like to emphasize that
spin sensitive bleaching is also expected for inter-subband
transitions in n-type QWs and may be used to extract spin
relaxation times of electrons in $n$-type materials.

We thank L. Golub for helpful discussions.
Financial support by the DFG, the RFBR, the INTAS and
the NATO linkage  program is gratefully acknowledge.

\newpage

%%%%%%%%%%%% Figure Captions %%%%%%%%%%%%%%%

\begin{figure}
\caption{
Sketch of
direct optical transitions (full line) between the first heavy hole and the first
light hole subband in p-GaAs/AlGaAs QWs. While the splitting of the bands in $k$-space is
necessary  for an understanding of the circular photogalvanic effect~\protect\cite{PRL01}
it is unimportant for the saturation process and 
ignored in the sketch. The absorption of far-infrared radiation with wavelength of 148~$\mu$m
 (photon energy $\hbar\omega=8.3$~meV) occurs
very close to $k = 0$. Therefore the initial and final states are characterized by
angular momentum quantum numbers $m = \pm 3/2$ and  $\pm 1/2$, respectively.
Dashed lines shows the energy
relaxation of photoexcited carriers. $\varepsilon_F$ is the Fermi energy.
}
\end{figure}

\begin{figure}
\caption{
Photogalvanic current $j_x$   normalized by the intensity $I$
as a function of $I$
for  circularly and linearly polarized radiation.
Measurements are presented for $T=~$20~K.
The  inset  shows   the geometry of the experiment where $\bbox{\hat{e}}$ 
indicates the direction of the incoming light.
The current $j_{x }$ flows along the
[1$\bar{1}$0]-
direction at normal
incidence of radiation on $p$-type (113)A- grown
GaAs/AlGaAs QWs.
In order to obtain the circular photogalvanic effect (CPGE) right or left circularly
polarized  light has been applied. To obtain the linear photogalvanic
effect (LPGE)
linearly polarized radiation with the electric field vector $E$
oriented at 45 degrees to the  direction $x$ was used.
The measurements  are
fitted
to $j_x/I\propto 1/(I+I/I_s)$ with one parameter $I_s$ for each state of polarization
(full line: circular, broken line: linear).
}
\end{figure}

\begin{figure}
\caption{Temperature dependence of the saturation intensity $I_s$ for
linearly and circularly polarized radiation. The dependence is shown for
one p-GaAs/AlGaAs (311)-MBE-grown sample with a single QW of $L_W$=15 nm width.
The
free carrier density is  $1.66\cdot 10^{11}$ cm$^{-2}$ and the mobility is 6.5x10$^5$~cm$^2$/(Vs).
Insets show a microscopic picture explaining the origin of the difference in
saturation intensities.
}
\end{figure}

\begin{figure}
\caption{Experimentally determined spin relaxation times $\tau_s$
of holes in $p$-type
GaAs/AlGaAs QWs as a function of temperature.
Open triangles and full dots corresponds to (113) MBE-grown
15~nm single and multiple (20) QWs, respectively.
Free carrier densities of all samples were about $2\cdot 10^{11}$ cm$^{-2}$
for each QW.
}
\end{figure}


\begin{thebibliography}{99}
\bibitem{photogalvanic} The photogalvanic currents $j_x$ generated by
the polarized radiation at normal incident has been measured
as a function of intensity in unbiased samples
via the voltage drop across a
50~$\Omega$ load resistor in a closed circuit configuration.
For LPGE
linearly polarized radiation with the electric field vector $E$
oriented at 45 degrees to the  direction $x$ was used.
\end{thebibliography}
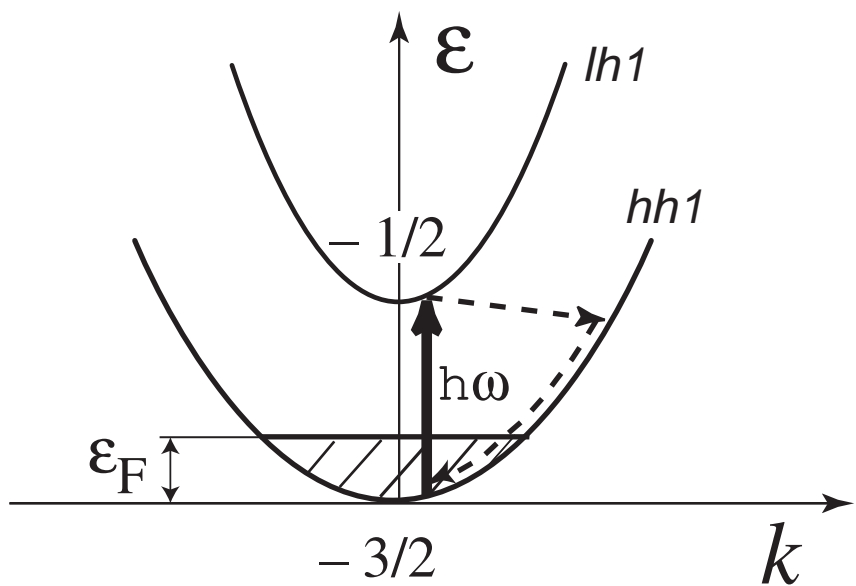
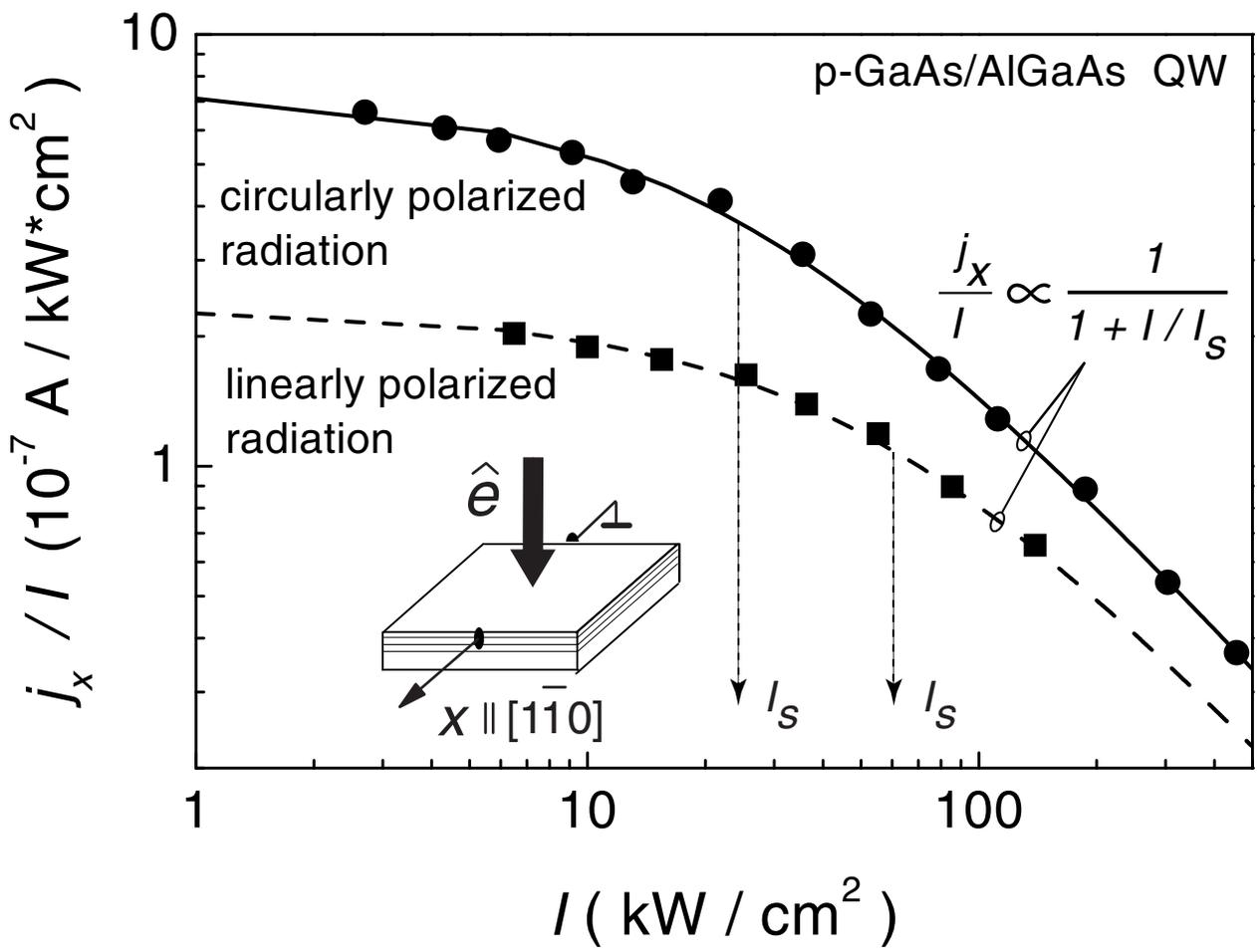
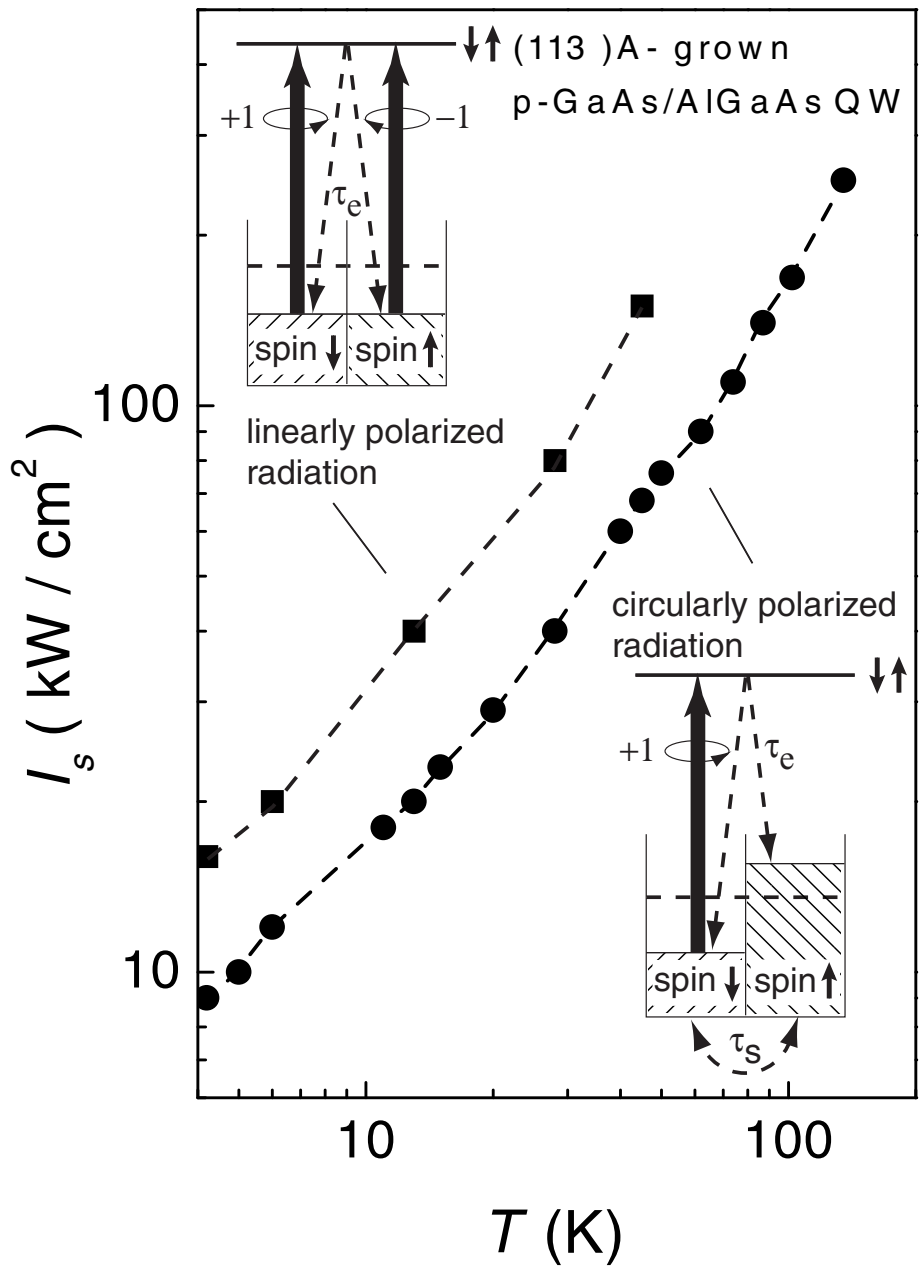
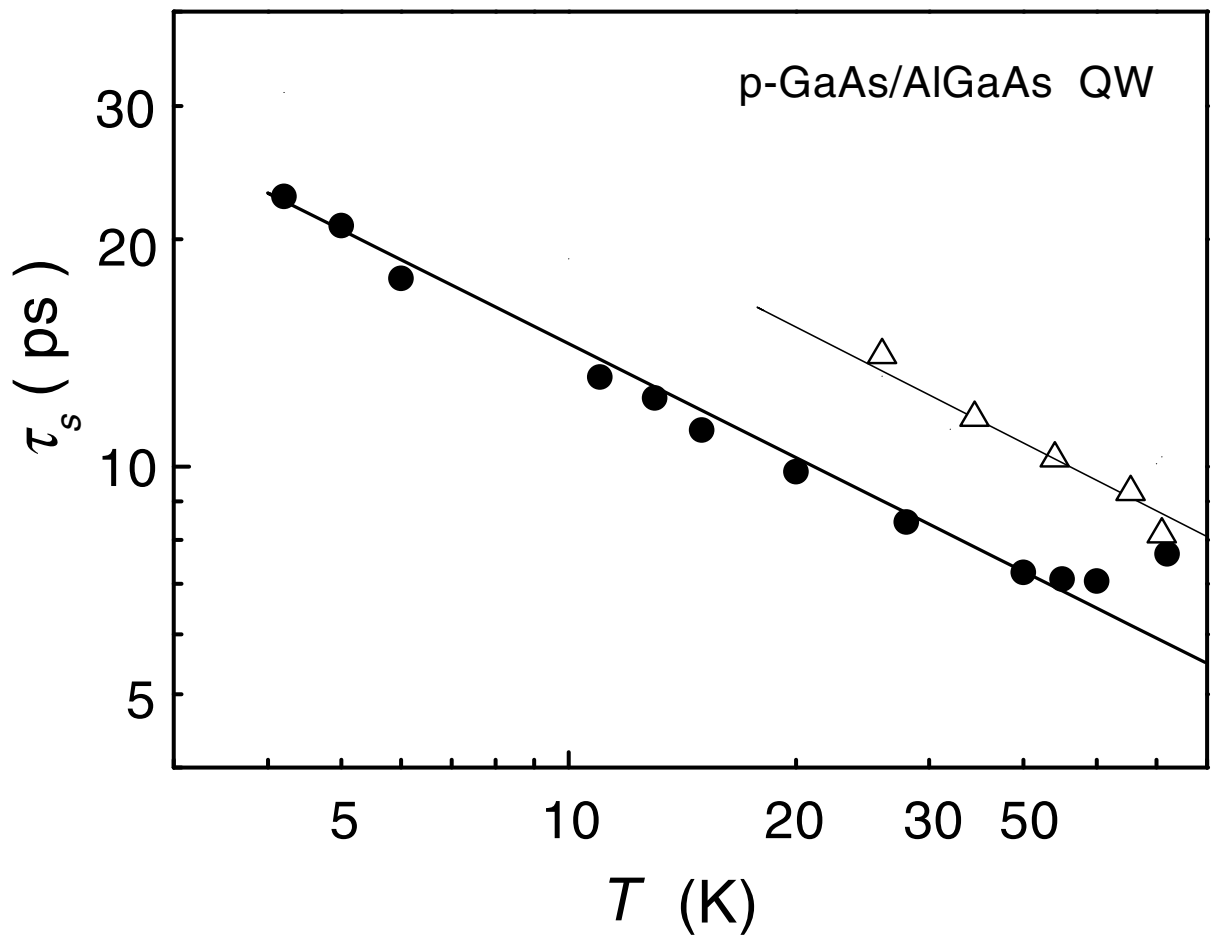
\end{document}